# A Comparative Study of Inter-Regional Intra-Industry Disparity

Samidh Pal[1] [2] [3]
ORCID ID: https://orcid.org/0000-0001-9227-6431

*Abstract:*
*This paper investigates the inter-regional intra-industry disparity within selected Indian manufacturing industries and industrial states. The study uses three measures - the Output-Capital Ratio, the Capital-Labor Ratio, and the Output-Labor Ratio - to critically evaluate the level of disparity in average efficiency of labor and capital, as well as capital intensity. Additionally, the paper compares the rate of disparity of per capita income between six major industrial states. The study finds that underutilization of capacity is driven by an unequal distribution of high-skilled labor supply and upgraded technologies. To address these disparities, the paper suggests that policymakers campaign for labor training and technology promotion schemes throughout all regions of India. By doing so, the study argues, the country can reduce regional inequality and improve economic outcomes for all.*

[1] *Pursuing Ph.D, Department of Economic Sciences, University of Warsaw, Poland. s.pal@uw.edu.pl;*

[2] **Statements and Declarations:** The authors of this research paper declare that they have no financial or personal relationships that could inappropriately influence or bias their work. They also have no professional or personal biases that could affect the research findings or interpretation of the results. The authors have no conflicts of interest to report.

[3] **Competing Interests:** The authors declare that they have no known competing financial interests or personal relationships that could have appeared to influence the work reported in this paper.

**Introduction**
The distribution and location of industrial activity across regions is a crucial aspect of economic development. Two theories that explain the spatial distribution of economic activity are the Growth Pole Theory and the Central Place Theory. The former posits that economic development is polarized and occurs through the emergence of growth poles with varying intensities, while the latter highlights the importance of factors such as transportation costs, demand function, and economies of scale in shaping the location of industrial plants. Through an analysis of the West Bengal region in India, this study explores the applicability of these theories and considers additional factors that may influence industrial location, such as access to raw materials, labor, and capital, as well as geographical factors like weather and landscapes. The study finds that the southern part of West Bengal, with its concentration of core industries, linked industries, and growth poles, is a hub of economic activity, while the northern part is dominated by agriculture and has limited industry. Overall, this study provides insights into the complex and multifaceted factors that shape the location of industrial activity and their implications for regional economic development.

**Study Background**
The spatial distribution of economic activities and inter-regional intra-industrial disparity have long been subjects of interest in economic geography. The Central Place Theory, first proposed by Walter Christaller in 1933 and further developed by Biermann (1973), explains the spatial distribution of cities and industries across the landscape based on factors such as transportation cost, demand function, and economies of scale. However, recent research suggests that the distribution of industries is not solely based on these factors but also on three ratio factors: Output-Capital Ratio, Output-Labor Ratio, and Capital-Labor Ratio.

In addition to the Central Place Theory, the Growth Pole Theory, proposed by Perroux, emphasizes that economic development is not uniform across a region, but rather specific poles or clusters of economic activities exist. Higgins and Savoie (2018) argue that imbalances between industry and geographical areas can hinder economic growth, particularly in regions with varying weather and landscapes. Furthermore, Slusarciuc (2015) notes that development does not appear everywhere at once, but rather occurs at specific points on growth poles with varying intensities.

Analyzing the industrial disparity in India, Myrdal (1957) identified a circular causation process that leads to the rapid development of highly developed regions, while weaker regions tend to remain poor and underdeveloped. The main causes of the backwardness of underdeveloped regions are the strong 'backwash effect' and the weak 'spread effect,' which determine the rate of growth of lag regions.[4] Gaile (1980) used the backwash effect to describe the potential negative effects of urban growth on peripheral areas.

The present study focuses on the inter-regional intra-industrial disparity among five major industrial states in India, compared with the state of West Bengal. The issue of regional disparity is a common phenomenon in India, which can be analyzed through inter-regional and intra-industry disparities. The time duration of industrial development in a particular region plays a crucial role in determining regional disparity (Williamson, 1965). Late industrial development in a region leads to an increase in disparity.

---

[4] The negative effect of the core's growth is on the periphery. Because of the out-migration of economically active people, outflows of capital, decreasing tax base the firms of the periphery are not able to compete with the firms of the core, and therefore periphery is being flooded with core's products.

The concept of beta convergence suggests that if a poor region grows faster than a rich region, then ultimately, the differences between regions with disparities will diminish (Solow, 1956). The neoclassical growth theory postulates that factors of production, primarily capital, are subject to diminishing returns to scale. This concept of diminishing returns is also suggested by Noorbakhsh (2006) for healthcare and education indicators. Thus, both low and high developed regions will converge over time. It is advantageous for both regions to have a development balance at a high level through the growth pole strategies.

However, in practice, there is a longer period of regional imbalances in the world than a balance at any level of development. Several factors, such as power supply, local skillful labor supply, transportation, supply of raw materials, etc., imply a development balance, which helps mitigate the unequal development of infrastructures and reduces inter-regional disparities. Consequently, industries will not only concentrate in a single region but will be spread all over the nation (O'Hara, 2008).

One of the characteristics of India is the centralization of capital (Chaudhuri et al., 2014). As a result, the concentration of output, capital stock, and employment are near urban areas (Amirapu et al., 2018). Therefore, urbanization is correlated with industrialization.

In the case of West Bengal, our study (see map: 1) shows that the concentration of industries is limited to the southern part of the region, where there is better access to raw materials and high-skilled labor. The emergence of secondary industries and linked industries in the northern part of West Bengal contributes to regional economic diversity.

Map: 1
Plots of industrial units of West Bengal coordinate wise through spatial data analysis.

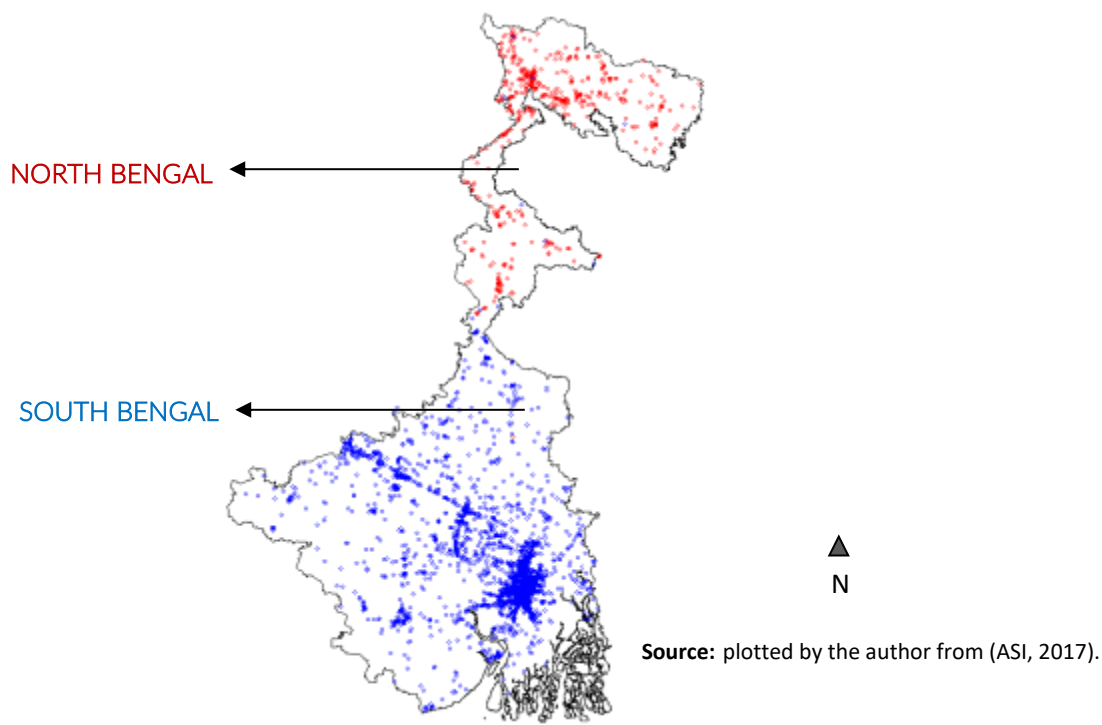

**Source:** plotted by the author from (ASI, 2017).

**Literature Review**
The literature reviewed above highlights the persistent issue of regional and industrial disparity in India. Nayak et al. (2010) found that the Gini coefficient has increased significantly from 0.164 to 0.245 between 1980 and 2007, indicating an increase in the disparity rate of national state domestic product with respect to per capita national income. Similarly, Sharma and Khosla (2013) examined inter-state disparities in the Indian industrial sector and identified a significant gap between states that has become more noticeable over time. Gradín (2018) investigated the sources of such variability in India and found that the composition of the workforce in each state was strongly associated with inequality gaps. Additionally, S.N. Nandy (2019) found socio-economic disparity among various states/regions/sectors in India, with the southern and north-western states performing better than the eastern and central parts of India, and the north-eastern states still lagging behind.

These findings highlight the need for a more detailed analysis of various sectors of the economy and various sections of the population to promote inclusive growth. To contribute to this understanding, the present research will focus on six manufacturing industrial states and conduct a comparative disparity study between Output-Capital Ratio, Capital-Labor Ratio, and Output-Labor ratio to identify overall disparities in industry group-wise. Additionally, we will compare the results of different indices to arrive at approximately accurate findings. We will also examine the disparity level of the per-capita income rate state-wise to provide a comprehensive analysis of regional and industrial disparities in India. Overall, this research will contribute to a better understanding of the sources and extent of socio-economic disparities in India, which can inform policy interventions to promote more inclusive growth.

**Data**
The study's data selection process involved utilizing the National Industrial Classification (NIC) system with three-digit codes to identify and collect data from all registered manufacturing industries in India. The Annual Survey of Industry (ASI) (Central Statistical Organization, 2013/2019) conducted by the Central Statistical Organization between 2010 to 2017 was the source of data for this research, including both private and public industries operating in the six major industrial states.

As per Pal (2019), we ranked the states according to their regional specialization coefficients, which ranked the most industrialized state to the least industrialized. This rank allowed us to measure the level of disparity in industries across the six states.

In line with the study's objective, we selected only those industries that were available in all six states and omitted others. We then calculated the capital-labor ratio, output-capital ratio, and output-labor ratio for each industry. To determine the ratios, we utilized value-added data as the output data, invested capital data as capital, and labor wage data as labor.

The results of the ratios' disparity between the six major industrial states were compared using four indices, namely the Gini coefficient, Ricci-Schutz coefficient, Atkinson measure, and Theli's index, to assess their similarity. This approach enabled us to investigate the extent of the disparities in the manufacturing industries' performance across the six states.

**Empirical Results and Interpretations**

We conducted an analysis of various economic indicators, including the Gini index, Ricci-Schutz coefficient, Atkinson's measure, and Theil's index, to measure disparities in the capital-labor ratio and output-labor ratio industry group-wise across selected states. Table 1 presents the highest and lowest levels of disparity in all indices for some common industrial groups. For a comprehensive overview of the results, please refer to *Appendix A*.

Table: 1

Comparative indices values of output-labor ratio and capital-labor ratio (Industry group wise)

| Code | Theil_APL | Gini_APL | Atkinson_APL | RS_APL | code | Theil_CLR | Gini_CLR | Atkinson_CLR | RS_CLR |
|---|---|---|---|---|---|---|---|---|---|
| **Higher Level of Disparity** | | | | | **Higher Level of Disparity** | | | | |
| *264 | NaN | 0.62 | NaN | 0.43 | 192 | 0.33 | 0.44 | 0.16 | 0.37 |
| 120 | 0.72 | 0.62 | 0.34 | 0.50 | 263 | 0.33 | 0.43 | 0.16 | 0.38 |
| 263 | 0.29 | 0.39 | 0.13 | 0.31 | *264 | 0.31 | 0.43 | 0.15 | 0.32 |
| 143 | 0.24 | 0.37 | 0.12 | 0.26 | 120 | 0.31 | 0.40 | 0.20 | 0.33 |
| 192 | 0.22 | 0.36 | 0.11 | 0.27 | 143 | 0.22 | 0.37 | 0.11 | 0.29 |
| **Lower Level of Disparity** | | | | | **Lower Level of Disparity** | | | | |
| 170 | 0.04 | 0.16 | 0.02 | 0.13 | 161 | 0.06 | 0.19 | 0.03 | 0.15 |
| 161 | 0.04 | 0.15 | 0.02 | 0.11 | 271 | 0.05 | 0.16 | 0.02 | 0.11 |
| 271 | 0.03 | 0.13 | 0.01 | 0.09 | 170 | 0.03 | 0.13 | 0.02 | 0.11 |
| 222 | 0.02 | 0.11 | 0.01 | 0.08 | 222 | 0.01 | 0.08 | 0.01 | 0.06 |
| *293 | 0.01 | 0.08 | 0.01 | 0.06 | *293 | 0.00 | 0.05 | 0.00 | 0.04 |

**Source:** Calculating by author from Annual Survey of Industries (ASI) data-2016-2017
Note: APL – Output-labor ratio and CLR – Capital-labor ratio

We used Lorenz curve analysis to demonstrate how the curves differ for each industry group at higher and lower levels of disparity. Figures 1 and 2 illustrate the disparity level of capital-labor ratio and output-labor ratio for two industry groups, namely, Manufacture of consumer electronics (NIC 264) and Manufacture of parts and accessories for motor vehicles (NIC 293). One is identified as having a higher level of disparity, while the other is identified as having a lower level of disparity.

Figure: 1

Lorenz curve analysis for industry group: (NIC 264) with capital-labor ratio and output-labor ratio

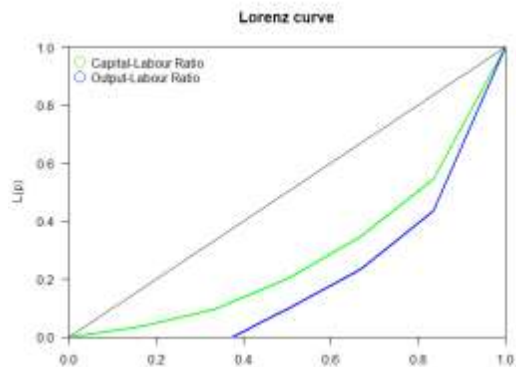

Source: Calculating by author from Annual Survey of Industries (ASI) data-2016-2017

Figure: 2

Lorenz curve analysis for industry group: (NIC 293) with capital-labor ratio and output-labor ratio

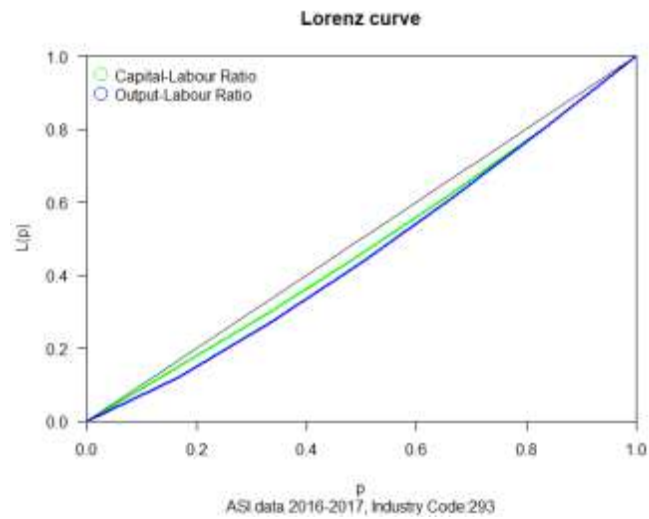

Source: Calculating by author from Annual Survey of Industries (ASI) data-2016-2017

Figures 3 and 4 compare the capital-labor ratio and output-labor ratio state-wise within the same industry group. For industry group NIC 264, the state-wise levels of capital-labor ratio and output-labor ratio are not equal, indicating underutilization of capacity in West Bengal. Conversely, for industry group NIC 293, the figures indicate that the levels of capital-labor ratio and output-labor ratio are almost equal state-wise.

Figure: 3

State wise comparative study of capital-labor ratio and output-labor ratio for industry group: (NIC 264)

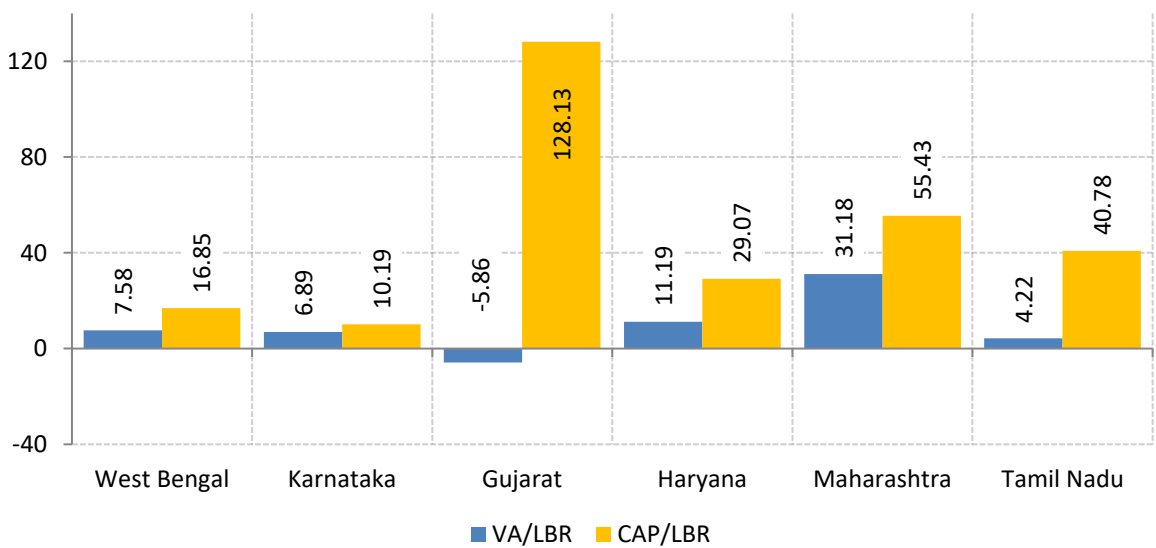

Source: Calculating by author from Annual Survey of Industries (ASI) data-2016-2017

Figure: 4
State wise comparative study of capital-labor ratio and output-labor ratio for industry group: (NIC 293)

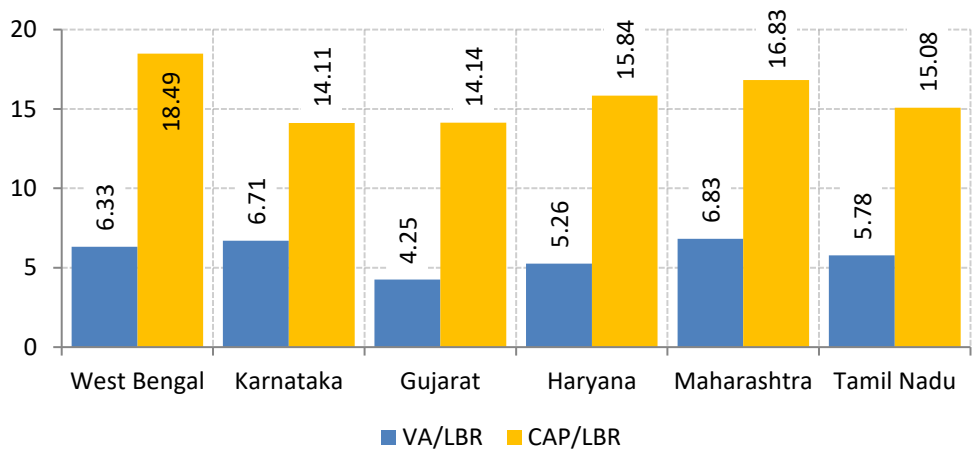

Source: Calculating by author from Annual Survey of Industries (ASI) data-2016-2017

In the second part of our interpretation, we classified the same industry groups into two sections based on higher and lower levels of disparity and observed that almost all industries within the same group ranked consistently higher or lower in terms of disparity levels across all four types of indices, including the output-capital ratio. The full results can be found in ***Appendix B***. We further examined two industry groups, NIC 264 and NIC 293, and depicted the disparity levels of their output-capital ratios using the Lorenz curve in Figure 5.

Table: 2
Comparative indices values of output-capital ratio (Industry group wise)

| code.1 | Theil_OCR | Gini_OCR | Atkinson_OCR | RS_OCR |
|---|---|---|---|---|
| **Higher Level of Disparity** | | | | |
| 192 | 0.48 | 0.49 | 0.21 | 0.42 |
| 120 | 0.32 | 0.43 | 0.17 | 0.33 |
| * 264 | NaN | 0.40 | NaN | 0.31 |
| 143 | 0.28 | 0.39 | 0.16 | 0.28 |
| 263 | 0.15 | 0.30 | 0.08 | 0.22 |
| **Lower Level of Disparity** | | | | |
| 275 | 0.08 | 0.22 | 0.04 | 0.18 |
| 170 | 0.04 | 0.16 | 0.02 | 0.12 |
| 222 | 0.03 | 0.15 | 0.02 | 0.11 |
| 271 | 0.02 | 0.10 | 0.01 | 0.07 |
| *293 | 0.01 | 0.08 | 0.01 | 0.06 |

**Source:** Calculating by author from Annual Survey of Industries (ASI) data-2016-2017. Note: OCR – Output-capital ratio

Figure: 5
Comparative Lorenz curve analysis for industry group: (NIC 264) and (NIC 293) with output-capital ratio

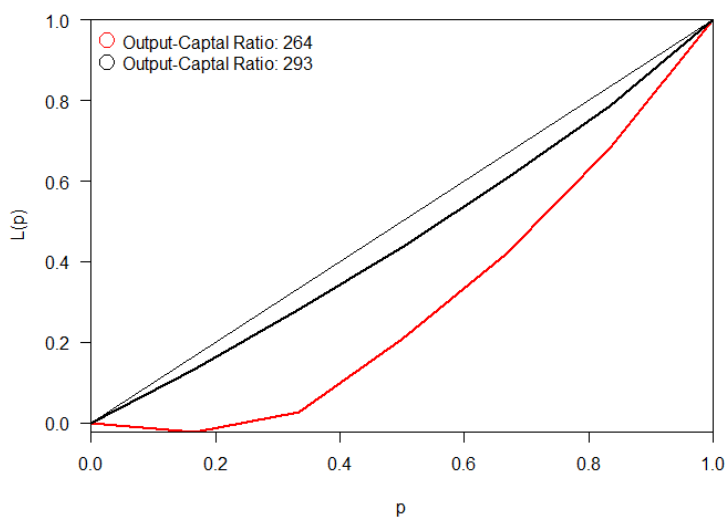

Source: Calculating by author from Annual Survey of Industries (ASI) data-2016-2017

We calculated the growth rate of output-capital ratio industry group-wise for every state from 2012 to 2017. Figure 6 illustrates that for industry group NIC 264, the growth rate of marginal efficiency of capital is lower in Gujarat, Tamil Nadu, and West Bengal compared to the other three states. Conversely, for industry group NIC 293, the growth rate of marginal efficiency of capital in West Bengal is higher than in the other states.

Figure: 6
State wise comparative growth rate of output-capital ratio for industry group: Manufacture of consumer electronics (NIC 264) and (NIC 293)

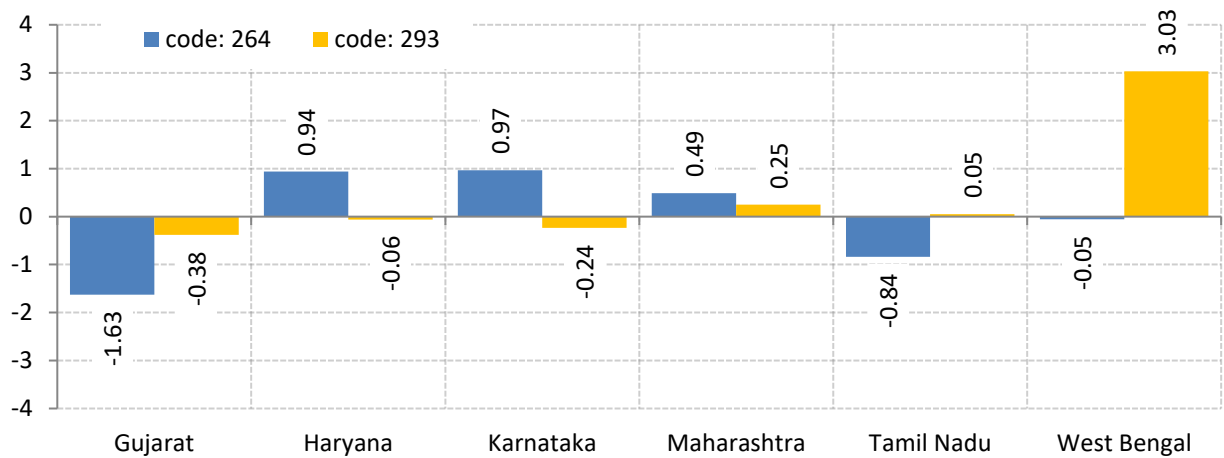

Source: Calculating by author from Annual Survey of Industries (ASI) data-2016-2017
Note: We are unable to take 2010-11 as a base year because Haryana had no data for industry group: 264

Moving on to the third part of our interpretation, we examined the rate of disparity in per-capita income between West Bengal and the other five selected states. Table 3 depicts the differences in per capita income between West Bengal and the other five states from 2010 to 2017. Using this small set of time series data, we plotted five trend lines on scattered plots. Table 4 shows the equation for each trend line, organized by state. The rates of per capita income disparity are as follows: 0.47 for Maharashtra-West Bengal, 0.37 for Karnataka-West Bengal, 0.53 for Haryana-West Bengal, 0.72 for Gujarat-West Bengal, and 0.15 for Tamil Nadu-West Bengal. The R-squared values indicate a good fit with the actual data, and all the linear trend lines show a continuous rise in per capita income disparity between West Bengal and the other five states from 2010 to 2017.

Table: 3

Yearly per capita income differences between West Bengal and five other states

| Years | GJ-WB | HR-WB | KA-WB | MH-WB | TM-WB |
|---|---|---|---|---|---|
| 2010-2011 | 3.41 | 1.15 | 1.90 | 5.22 | 0.53 |
| 2011-2012 | 2.58 | 1.57 | 1.70 | 4.30 | 0.62 |
| 2012-2013 | 4.92 | 2.92 | 2.87 | 6.14 | 1.30 |
| 2013-2014 | 5.20 | 2.57 | 2.56 | 7.00 | 0.92 |
| 2014-2015 | 7.40 | 3.10 | 2.41 | 7.30 | 0.73 |
| 2015-2016 | 7.47 | 3.69 | 3.09 | 7.28 | 1.34 |
| 2016-2017 | 6.06 | 4.61 | 4.56 | 7.26 | 1.64 |

Source: Calculating by author from Annual Survey of Industries (ASI) data-2010-2017

Table: 4

Linear trend line equations from differences between West Bengal and five other states

| Differences | Linear trend line equations |
|---|---|
| Maharashtra-West Bengal | y = **0.47**x + 4.47, $R^2$ = 0.73 |
| Karnataka- West Bengal | y = **0.37**x + 1.26, $R^2$ = 0.70 |
| Haryana-West Bengal | y = **0.53**x + 0.69, $R^2$ = 0.92 |
| Gujarat-West Bengal | y = **0.72**x + 2.40, $R^2$ = 0.70 |
| Tamil Nadu-West Bengal | y = **0.15**x - 0.40, $R^2$ = 0.60 |

Source: Calculating by author from Annual Survey of Industries (ASI) data-2010-2017

Table 5 provides an overview of the yearly state-wise rank of average efficiency of labor, capital intensity, and average efficiency of capital. It is apparent that West Bengal consistently ranks lowest in output-labor ratio and output-capital ratio compared to other states since 2010. While West Bengal has the third-highest level of capital-labor ratio in 2017, indicating an improvement in capital intensity across sectors, underutilization still persists.

Table: 5

Comparative efficiency ratios of manufacturing industries: 2010-11 to 2016-17

| Position | 1 | 2 | 3 | 4 | 5 | 6 |
|---|---|---|---|---|---|---|
| 2010-11 | Maharashtra | Gujarat | Karnataka | Haryana | Tamil Nadu | West Bengal |
| O/L | 8.815367 | 6.905383 | 5.227061 | 4.517775 | 3.704652 | 3.281758 |
| 2011-12 | Maharashtra | Gujarat | Haryana | Karnataka | Tamil Nadu | West Bengal |
| O/L | 8.335908 | 6.337112 | 5.126102 | 5.09702 | 3.965115 | 3.142148 |
| 2012-13 | Maharashtra | Gujarat | Haryana | Karnataka | Tamil Nadu | West Bengal |
| O/L | 10.01336 | 8.71763 | 6.853529 | 6.202641 | 4.649544 | 3.395782 |
| 2013-14 | Maharashtra | Gujarat | Haryana | Karnataka | Tamil Nadu | West Bengal |
| O/L | 10.6317 | 9.3555 | 6.057261 | 5.966577 | 4.261443 | 3.474723 |
| 2014-15 | Gujarat | Maharashtra | Haryana | Karnataka | Tamil Nadu | West Bengal |
| O/L | 11.60357 | 11.08635 | 6.55833 | 5.85159 | 4.129514 | 2.913387 |
| 2015-16 | Gujarat | Maharashtra | Haryana | Karnataka | Tamil Nadu | West Bengal |
| O/L | 11.51022 | 11.0258 | 7.093119 | 6.394097 | 4.685265 | 3.549956 |
| 2016-17 | Gujarat | Haryana | Karnataka | Maharashtra | Tamil Nadu | West Bengal |
| O/L | 10.12093 | 8.090441 | 7.771315 | 10.84026 | 4.895027 | 4.533546 |
| 2010-11 | Maharashtra | Tamil Nadu | Haryana | Karnataka | Gujarat | West Bengal |
| O/K | 0.42199 | 0.309562 | 0.284866 | 0.266287 | 0.22384 | 0.22359 |
| 2011-12 | Maharashtra | Haryana | Tamil Nadu | Karnataka | Gujarat | West Bengal |
| O/K | 0.38015 | 0.323394 | 0.296349 | 0.250189 | 0.191864 | 0.188165 |
| 2012-13 | Maharashtra | Haryana | Tamil Nadu | Karnataka | Gujarat | West Bengal |
| O/K | 0.359448 | 0.329906 | 0.312614 | 0.257651 | 0.243363 | 0.181208 |
| 2013-14 | Maharashtra | Haryana | Karnataka | Gujarat | Tamil Nadu | West Bengal |
| O/K | 0.411597 | 0.329818 | 0.237076 | 0.234287 | 0.229185 | 0.181096 |
| 2014-15 | Maharashtra | Haryana | Gujarat | Tamil Nadu | Karnataka | West Bengal |
| O/K | 0.419356 | 0.359047 | 0.279722 | 0.266927 | 0.243966 | 0.139449 |
| 2015-16 | Maharashtra | Haryana | Tamil Nadu | Karnataka | Gujarat | West Bengal |
| O/K | 0.40936 | 0.335873 | 0.311415 | 0.268509 | 0.262041 | 0.169132 |
| 2016-17 | Maharashtra | Haryana | Karnataka | Tamil Nadu | Gujarat | West Bengal |
| O/K | 0.415273 | 0.370984 | 0.328595 | 0.292842 | 0.2031 | 0.182144 |
| 2010-11 | Gujarat | Maharashtra | Karnataka | Haryana | West Bengal | Tamil Nadu |
| K/L | 30.84962 | 20.88997 | 19.62941 | 15.85932 | 14.67755 | 11.96738 |
| 2011-12 | Gujarat | Maharashtra | Karnataka | West Bengal | Haryana | Tamil Nadu |
| K/L | 33.02914 | 21.92793 | 20.37264 | 16.69892 | 15.85093 | 13.37989 |
| 2012-13 | Gujarat | Maharashtra | Karnataka | Haryana | West Bengal | Tamil Nadu |
| K/L | 35.82158 | 27.85759 | 24.07377 | 20.77418 | 18.73968 | 14.87313 |
| 2013-14 | Gujarat | Maharashtra | Karnataka | West Bengal | Tamil Nadu | Haryana |
| K/L | 39.93187 | 25.83034 | 25.16731 | 19.18717 | 18.59388 | 18.36548 |
| 2014-15 | Gujarat | Maharashtra | Karnataka | West Bengal | Haryana | Tamil Nadu |
| K/L | 41.48245 | 26.43663 | 23.9853 | 20.89215 | 18.26592 | 15.47059 |
| 2015-16 | Gujarat | Maharashtra | Karnataka | Haryana | West Bengal | Tamil Nadu |
| K/L | 43.92528 | 26.93425 | 23.81336 | 21.11845 | 20.98924 | 15.0451 |
| 2016-17 | Gujarat | Maharashtra | West Bengal | Karnataka | Haryana | Tamil Nadu |
| K/L | 49.83224 | 26.10396 | 24.88983 | 23.6501 | 21.80807 | 16.71561 |

**Source:** Calculating by author from Annual Survey of Industries (ASI) data-2010-2011 to2016-2017

Furthermore, Figure 7 illustrates the annual per-capita profit share within the six major industrial states, indicating that West Bengal has consistently had the smallest profit share since 2010 compared to the other five states.

Figure: 7
Per-capita profit share within six different states 2010-2017 (Rs. in millions)

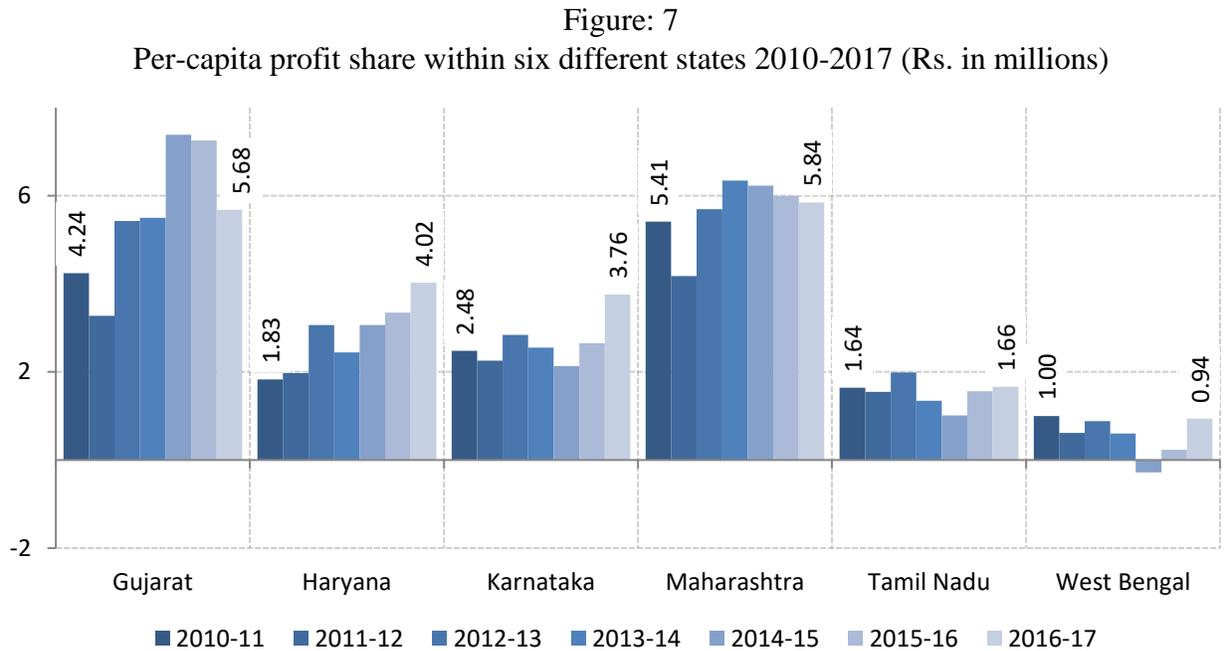

**Source:** Calculating by author from Annual Survey of Industries (ASI) data-2010-2011 to 2016-2017

Note: 1 lakh = 0.1 million

**Conclusion**

Based on our findings, it appears that the manufacture of consumer electronics (NIC 264) in West Bengal has lower capital intensity and lower average labor efficiency compared to other states in India. This suggests that the region is not utilizing its capacity efficiently in this industry, resulting in an inter-regional intra-industry disparity in India, particularly in West Bengal. The poor performance of industrial activity in the region creates a barrier against new firms with advanced technologies, thereby keeping the region less industrialized.

To identify the primary causes of excess capacity industry-wise, we will investigate the responses of capital, labor, and capital intensity. Myrdal's circular and cumulative causation theory consists of three main Kaldor's laws, which discuss the effects of increasing returns in the manufacturing sector on macroeconomic dynamics. According to the first law, the growth rate of manufacturing production is positively related to GDP, and increasing returns-to-scale prevails in the manufacturing sector. These have dynamic and macroeconomic effects, including "learning by doing" and technological innovation.

The unequal distribution of high-skilled labor supply is due to the unequal distribution of per capita income and the unequal distribution of upgraded machines. If workers are not upgraded by using advanced technology, they are not offered high-paying jobs (Gradín, 2018). Advanced technology always encourages producing more efficient output, leading to more profits, revenue, and higher wages for labor. One of the reasons behind the migration of workers to other states is job uncertainty, a reduction in livelihood opportunities, and low growth rates in agro-dominated rural areas (Mahapatro, 2013).

**Policies and Suggestion**

Various policies and schemes have been introduced in India to reduce the differences among the six major industrial states, with the central and state governments implementing various measures. One such scheme is the Pradhan Mantri Kaushal Vikas Yojna (PMKVY), a grant-based initiative launched on 2 October 2016 with the aim of providing free training and certification for skill development in more than 252 job roles, enhancing employability among school and college dropouts and unemployed youth, and improving the quality of training infrastructure. The PMKVY also seeks to encourage standardization in the certification process and create a registry of skills (Ministry of Skill Development and Entrepreneurship, 2016-20).

In addition, the Credit Linked Capital Subsidy Scheme (CLCSS) has been introduced by the Development Commissioner, MSME (Ministry of Micro, Small and Medium Enterprises), to support manufacturing industries. The CLCSS aims to provide an upfront capital subsidy to upgrade plants and machinery for existing and new manufacturing units that adopt eligible and proven technology approved under the scheme guidelines. The objective of the CLCSS is to promote the adoption of upgraded technologies and provide import subsidies, promote zero-defect and zero-effect practices in manufacturing processes, and encourage the adoption of new global standard technologies (Ministry of Micro, Small and Medium Enterprises, 2006).

However, these policies and schemes are not evenly promoted throughout the country, and many regions remain unaware of their benefits. To address this issue, policymakers should focus on promoting and redesigning labor training schemes. To promote schemes regionally and industry-wise, media attention such as television, banners, posters, newspapers, etc., should be used. This redevelopment process can encourage existing entrepreneurs to expand their units to different regions, while new entrepreneurs in backward regions can be motivated to establish new units, leading to job creation. This expansion and job creation will increase the demand for training camps. The training schemes should be redesigned to include not only short-term but also long-term courses and refurbished with advanced and quality training infrastructure. The successful reimplementation of these schemes requires greater involvement from state governments and the central government.

In conclusion, these policies and schemes have been introduced to address the inter-regional intra-industry disparity in India. However, more efforts are needed to promote them equally across the country. The suggestion is to focus on the use of promotion and redesign of labor training schemes and to involve all state governments with the central government in the process to achieve successful reimplementation.

**References:**
Amirapu, A., Hasan, R., Jiang, Y., & Klein, A. (2018). *Geographic Concentration in Indian Manufacturing and Service Industries: Evidence from 1998 to 2013*. Asian Economic Policy Review, 14(1), 148–168. https://doi.org/10.1111/aepr.12251
Biermann, H. (1973). *A Note on the Central Place Theory*. Environment and Planning A: Economy and Space, 5(5), 649–654. https://doi.org/10.1068/a050649
Chaudhuri, B., Sarkar, S. and Panigrahi, A. K.(2014). *Chronic problems in industrialization in West Bengal*. International Growth Centre (IGC), 2015(01). https://www.theigc.org/wp-content/uploads/2015/01/Chaudhuri-et-al-2014.pdf
Gaile, G. L. (1980) *The Spread-Backwash Concept*. Regional Studies, (14) 15-25.
Gradín C. (2018) *Explaining cross-state earnings inequality differentials in India- An RIF decomposition approach* [Online]. WIDER Working Paper 24/2018.
Higgins B. and Savoie D. J. (2018) *Regional Economic Development – Essay in Honour of Francois Perroux*. Routledge.
Kaldor, N. (1966). *Causes of the Slow Rate of Economic Growth of the United Kingdom*. An Inaugural Lecture. London, Cambridge University Press, Recherches Economiques De Louvain / Louvain Economic Review, 34(2), 222-222. https://doi.org/10.1017/S0770451800040616
Mahapatra, S.R. (2012, June 13–16). *The changing pattern of internal migration in India: issues and challenges* [Paper presentation]. European Population Conference, Stockholm, Sweden. https://epc2012.princeton.edu/sessions/44
Ministry of Micro, Small & Medium Enterprises (2006). *Credit Linked Capital Subsidy Scheme (CLCSS) for Technological Upgradation*. Government of India. http://www.dcmsme.gov.in/schemes/Credit_link_Scheme.htm
Ministry of Skill Development & Entrepreneurship (2016-20). *Pradhan Mantri Kaushal Vikas Yojna (PMKVY) - Guidelines*. Government of India. http://pmkvyofficial.org/App_Documents/News/PMKVY%20Guidelines%20(2016-2020).pdf
Myrdal, G. (2021). *Economic Theory and Under-Developed Regions* [First edition]. Gerald Duckworth & Co.
Nandy, S. N. (2019). *Development Disparities in India: An Inter-State and Intra-State Comparison*. Journal of Land and Rural Studies, 7(2), 99–120. https://doi.org/10.1177/2321024919844407
Nayak, P. K., Chattopadhyay, S. K., Kumar, A. V. and Dhanya, V. (2010) *Inclusive Growth and its Regional Dimension* [Occasional Paper]. Reserve Bank of India.31(3). https://www.rbi.org.in/scripts/bs_viewcontent.aspx?Id=2359
Noorbakhsh, F. (2006) *International Convergence or Higher Inequality in Human Development? Evidence for 1975 to 2003* [Online]. UNU-WIDER Working Paper 15/2006. https://www.wider.unu.edu/sites/default/files/rp2006-15.pdf
O'Hara, P. A. (2008). *Principle of Circular and Cumulative Causation: Fusing Myrdalian and Kaldorian Growth and Development Dynamics*. Journal of Economic Issues, 42(2), 375–387. https://doi.org/10.1080/00213624.2008.11507146
Pal, S. (2019). *Measuring the Industrial Concentration and Regional Specialisation of Major Indian Industrial States*. The Indian Economic Journal, 67(3–4), 216–232. https://doi.org/10.1177/0019466220946330
Sharma, M., & Khosla, R. (2013). *Regional Disparities in India's Industrial Development: Discriminant Function Approach*. Indian Journal of Industrial Relations, 48(4), 692-702. Retrieved July 16, 2021, from http://www-1jstor-1org-1000094ro0382.han.buw.uw.edu.pl/stable/23509824
Slusarciuc, M. (2015). *Theories of Development Poles Applicability in the European Union Neighbourhood Cross-Border Frame*. Studia Ubb Negotia, 1, 41-57.

https://www.researchgate.net/publication/328466448_THEORIES_OF_DEVELOPMENT_POLES_APPLICABILITY_IN_THE_EUROPEAN_UNION_NEIGHBOURHOOD_CROSS-BORDER_FRAME

Solow, R. M. (1956). *A Contribution to the Theory of Economic Growth.* The Quarterly Journal of Economics, 70(1), 65. https://doi.org/10.2307/1884513

Vietorisz, T. and Harrison B. (1973) Labor Market Segmentation: Positive Feedback and Divergent Development. American Economic Review, 63, 366-376.

Williamson, J. G. (1965). *Regional Inequality and the Process of National Development: A Description of the Patterns.* Economic Development and Cultural Change, 13(4, Part 2), 1–84. https://doi.org/10.1086/450136

# Appendix: A

Comparative indices values of output-labor ratio and capital-labor ratio (Industry group wise)

| code | Theil_APL | Gini_APL | Atkinson_APL | RS_APL | code | Theil_CLR | Gini_CLR | Atkinson_CLR | RS_CLR |
|---|---|---|---|---|---|---|---|---|---|
| 264 | NaN | 0.62 | NaN | 0.43 | 301 | 0.69 | 0.58 | 0.31 | 0.51 |
| 120 | 0.72 | 0.62 | 0.34 | 0.50 | 192 | 0.33 | 0.44 | 0.16 | 0.37 |
| 291 | NaN | 0.50 | NaN | 0.35 | 263 | 0.33 | 0.43 | 0.16 | 0.38 |
| 104 | 0.29 | 0.40 | 0.15 | 0.34 | 264 | 0.31 | 0.43 | 0.15 | 0.32 |
| 275 | 0.27 | 0.40 | 0.14 | 0.32 | 120 | 0.31 | 0.40 | 0.20 | 0.33 |
| 321 | 0.29 | 0.39 | 0.13 | 0.31 | 279 | 0.33 | 0.39 | 0.14 | 0.35 |
| 263 | 0.29 | 0.39 | 0.13 | 0.31 | 143 | 0.22 | 0.37 | 0.11 | 0.29 |
| 143 | 0.24 | 0.37 | 0.12 | 0.26 | 302 | 0.12 | 0.27 | 0.06 | 0.22 |
| 192 | 0.22 | 0.36 | 0.11 | 0.27 | 292 | 0.11 | 0.25 | 0.05 | 0.18 |
| 139 | 0.23 | 0.35 | 0.11 | 0.28 | 275 | 0.11 | 0.25 | 0.05 | 0.19 |
| 309 | 0.27 | 0.35 | 0.19 | 0.27 | 201 | 0.10 | 0.25 | 0.05 | 0.21 |
| 302 | 0.18 | 0.33 | 0.09 | 0.24 | 331 | 0.11 | 0.25 | 0.05 | 0.20 |
| 241 | 0.22 | 0.32 | 0.15 | 0.22 | 274 | 0.11 | 0.25 | 0.05 | 0.22 |
| 331 | 0.16 | 0.30 | 0.08 | 0.26 | 139 | 0.11 | 0.25 | 0.05 | 0.19 |
| 272 | 0.13 | 0.28 | 0.06 | 0.21 | 239 | 0.13 | 0.25 | 0.07 | 0.16 |
| 151 | 0.13 | 0.28 | 0.06 | 0.21 | 202 | 0.12 | 0.25 | 0.07 | 0.20 |
| 108 | NaN | 0.27 | NaN | 0.21 | 242 | 0.11 | 0.25 | 0.05 | 0.18 |
| 581 | 0.11 | 0.26 | 0.06 | 0.20 | 141 | 0.10 | 0.24 | 0.05 | 0.20 |
| 107 | 0.11 | 0.26 | 0.06 | 0.22 | 105 | 0.10 | 0.24 | 0.05 | 0.20 |
| 201 | 0.11 | 0.26 | 0.06 | 0.18 | 221 | 0.09 | 0.23 | 0.04 | 0.18 |
| 242 | 0.11 | 0.26 | 0.05 | 0.19 | 241 | 0.08 | 0.22 | 0.04 | 0.16 |
| 110 | 0.12 | 0.25 | 0.06 | 0.20 | 107 | 0.08 | 0.22 | 0.04 | 0.18 |
| 202 | 0.12 | 0.25 | 0.06 | 0.17 | 321 | 0.10 | 0.22 | 0.04 | 0.17 |
| 265 | 0.10 | 0.24 | 0.05 | 0.18 | 310 | 0.08 | 0.22 | 0.04 | 0.17 |
| 274 | 0.11 | 0.24 | 0.05 | 0.17 | 291 | 0.10 | 0.21 | 0.05 | 0.19 |
| 106 | 0.09 | 0.23 | 0.05 | 0.16 | 273 | 0.08 | 0.21 | 0.04 | 0.15 |
| 162 | 0.08 | 0.23 | 0.04 | 0.17 | 210 | 0.08 | 0.21 | 0.05 | 0.17 |
| 281 | 0.09 | 0.22 | 0.05 | 0.16 | 151 | 0.07 | 0.21 | 0.03 | 0.17 |
| 282 | 0.08 | 0.22 | 0.04 | 0.16 | 259 | 0.07 | 0.20 | 0.03 | 0.17 |
| 239 | 0.08 | 0.22 | 0.04 | 0.16 | 265 | 0.08 | 0.20 | 0.04 | 0.19 |
| 243 | 0.07 | 0.21 | 0.04 | 0.15 | 261 | 0.07 | 0.20 | 0.03 | 0.14 |
| 103 | 0.07 | 0.20 | 0.04 | 0.14 | 104 | 0.07 | 0.20 | 0.03 | 0.14 |
| 329 | 0.07 | 0.20 | 0.03 | 0.15 | 161 | 0.06 | 0.19 | 0.03 | 0.15 |
| 325 | 0.06 | 0.19 | 0.03 | 0.15 | 106 | 0.06 | 0.19 | 0.03 | 0.14 |
| 181 | 0.06 | 0.19 | 0.03 | 0.16 | 309 | 0.06 | 0.19 | 0.03 | 0.14 |
| 310 | 0.06 | 0.19 | 0.03 | 0.15 | 231 | 0.05 | 0.18 | 0.03 | 0.15 |
| 221 | 0.07 | 0.18 | 0.04 | 0.14 | 108 | 0.05 | 0.18 | 0.03 | 0.15 |
| 231 | 0.05 | 0.18 | 0.03 | 0.13 | 131 | 0.06 | 0.18 | 0.04 | 0.12 |

| 279 | 0.05 | 0.18 | 0.03 | 0.14 | 251 | 0.06 | 0.17 | 0.03 | 0.14 |
| --- | --- | --- | --- | --- | --- | --- | --- | --- | --- |
| 210 | 0.05 | 0.17 | 0.02 | 0.14 | 181 | 0.06 | 0.17 | 0.03 | 0.14 |
| 152 | 0.05 | 0.17 | 0.03 | 0.12 | 152 | 0.06 | 0.17 | 0.03 | 0.13 |
| 170 | 0.04 | 0.16 | 0.02 | 0.13 | 271 | 0.05 | 0.16 | 0.02 | 0.11 |
| 261 | 0.04 | 0.16 | 0.02 | 0.11 | 162 | 0.04 | 0.16 | 0.02 | 0.13 |
| 141 | 0.04 | 0.15 | 0.02 | 0.13 | 281 | 0.05 | 0.16 | 0.02 | 0.12 |
| 161 | 0.04 | 0.15 | 0.02 | 0.11 | 329 | 0.04 | 0.16 | 0.02 | 0.12 |
| 273 | 0.04 | 0.15 | 0.02 | 0.13 | 110 | 0.04 | 0.16 | 0.02 | 0.11 |
| 292 | 0.03 | 0.13 | 0.01 | 0.10 | 325 | 0.04 | 0.15 | 0.02 | 0.12 |
| 271 | 0.03 | 0.13 | 0.01 | 0.09 | 103 | 0.03 | 0.13 | 0.02 | 0.10 |
| 222 | 0.02 | 0.11 | 0.01 | 0.08 | 170 | 0.03 | 0.13 | 0.02 | 0.11 |
| 251 | 0.02 | 0.11 | 0.01 | 0.10 | 272 | 0.03 | 0.12 | 0.01 | 0.11 |
| 131 | 0.02 | 0.11 | 0.01 | 0.08 | 243 | 0.02 | 0.10 | 0.01 | 0.07 |
| 259 | 0.02 | 0.10 | 0.01 | 0.08 | 222 | 0.01 | 0.08 | 0.01 | 0.06 |
| 293 | 0.01 | 0.08 | 0.01 | 0.06 | 581 | 0.01 | 0.08 | 0.01 | 0.06 |
| 105 | 0.01 | 0.07 | 0.00 | 0.06 | 282 | 0.01 | 0.07 | 0.01 | 0.05 |
| 301 | NaN | -1.45 | NaN | -1.18 | 293 | 0.00 | 0.05 | 0.00 | 0.04 |

**Source:** Calculating by author from Annual Survey of Industries (ASI) data-2016-2017
Note: APL – Output-labor ratio and CLR – Capital-labor ratio

# Appendix: B

Comparative indices values of output-capital ratio (Industry group wise)

| code | Theil_OCR | Gini_OCR | Atkinson_OCR | RS_OCR | code | Theil_OCR | Gini_OCR | Atkinson_OCR | RS_OCR |
|---|---|---|---|---|---|---|---|---|---|
| **301** | NaN | 1.05 | NaN | 0.81 | **265** | 0.06 | 0.20 | 0.03 | 0.15 |
| **291** | NaN | 0.55 | NaN | 0.39 | **239** | 0.06 | 0.19 | 0.03 | 0.14 |
| **192** | 0.48 | 0.49 | 0.21 | 0.42 | **329** | 0.07 | 0.19 | 0.03 | 0.16 |
| **302** | 0.47 | 0.48 | 0.21 | 0.42 | **251** | 0.06 | 0.19 | 0.03 | 0.16 |
| **120** | 0.32 | 0.43 | 0.17 | 0.33 | **151** | 0.06 | 0.19 | 0.03 | 0.14 |
| **264** | NaN | 0.40 | NaN | 0.31 | **221** | 0.06 | 0.18 | 0.03 | 0.14 |
| **143** | 0.28 | 0.39 | 0.16 | 0.28 | **282** | 0.06 | 0.18 | 0.03 | 0.14 |
| **241** | 0.27 | 0.36 | 0.17 | 0.24 | **103** | 0.08 | 0.18 | 0.05 | 0.15 |
| **104** | 0.19 | 0.33 | 0.10 | 0.24 | **105** | 0.06 | 0.18 | 0.03 | 0.15 |
| **309** | 0.24 | 0.33 | 0.17 | 0.25 | **325** | 0.06 | 0.18 | 0.03 | 0.12 |
| **108** | NaN | 0.32 | NaN | 0.27 | **331** | 0.06 | 0.18 | 0.03 | 0.13 |
| **263** | 0.15 | 0.30 | 0.08 | 0.22 | **141** | 0.05 | 0.17 | 0.03 | 0.16 |
| **272** | 0.15 | 0.28 | 0.07 | 0.22 | **131** | 0.06 | 0.17 | 0.03 | 0.14 |
| **273** | 0.13 | 0.28 | 0.07 | 0.20 | **202** | 0.05 | 0.17 | 0.02 | 0.14 |
| **162** | 0.13 | 0.27 | 0.06 | 0.20 | **259** | 0.05 | 0.17 | 0.03 | 0.12 |
| **110** | 0.11 | 0.25 | 0.05 | 0.18 | **170** | 0.04 | 0.16 | 0.02 | 0.12 |
| **279** | 0.11 | 0.25 | 0.06 | 0.18 | **261** | 0.07 | 0.15 | 0.04 | 0.12 |
| **581** | 0.11 | 0.24 | 0.06 | 0.19 | **181** | 0.04 | 0.15 | 0.02 | 0.12 |
| **292** | 0.09 | 0.24 | 0.05 | 0.19 | **222** | 0.03 | 0.15 | 0.02 | 0.11 |
| **242** | 0.10 | 0.24 | 0.05 | 0.18 | **210** | 0.03 | 0.14 | 0.01 | 0.10 |
| **310** | 0.10 | 0.23 | 0.05 | 0.18 | **139** | 0.03 | 0.13 | 0.01 | 0.11 |
| **201** | 0.09 | 0.23 | 0.04 | 0.18 | **274** | 0.03 | 0.13 | 0.02 | 0.10 |
| **161** | 0.09 | 0.23 | 0.04 | 0.18 | **231** | 0.02 | 0.12 | 0.01 | 0.09 |
| **281** | 0.09 | 0.23 | 0.04 | 0.20 | **107** | 0.02 | 0.11 | 0.01 | 0.08 |
| **243** | 0.08 | 0.22 | 0.04 | 0.17 | **152** | 0.02 | 0.10 | 0.01 | 0.09 |
| **275** | 0.08 | 0.22 | 0.04 | 0.18 | **271** | 0.02 | 0.10 | 0.01 | 0.07 |
| **106** | 0.11 | 0.22 | 0.05 | 0.19 | **293** | 0.01 | 0.08 | 0.01 | 0.06 |
| **321** | 0.06 | 0.20 | 0.03 | 0.16 | | | Note: OCR – Output-capital ratio | | |

**Source:** Calculating by author from Annual Survey of Industries (ASI) data-2016-2017